\documentclass{article}
\usepackage{LaThuileFPSpro}
\def\lsim{\lower.7ex\hbox{${\buildrel < \over \sim}$}}
\def\gsim{\lower.7ex\hbox{${\buildrel > \over \sim}$}}
\begin{document}
\title{
AIR-SHOWERS IN SPACE  AND Z-SHOWERS IN UNIVERSE FOR NEUTRINO ASTRONOMY AND SPECTROSCOPY}
\author{
  D.Fargion \\
  {\em Physics Department, Rome University1, La Sapienza, Rome, Italy}\\
  {\em  INFN,Rome1,Italy}\\
  E-mail address:~{\tt daniele.fargion@roma1.infn.it}\\
  } \maketitle

\baselineskip=10.0pt

\begin{abstract}
Amplified Tau-airshower at horizons may well open a novel powerful
windows, at PeV-EeV energy, to Neutrino Astronomy.
 Neutrino induced air-showering astronomy rise because of neutrino masses, their mixing and the
consequent replenishment of tau flavor during neutrino flight into
spaces; Tau-Air-Showers escaping the Earth are the main traces
amplified by its millions muon, billions gamma and thousand
billions photon secondaries. Earth edges and its sharp shadows is
the huge beam-dump detector for UHE neutrino and the almost
noise-free screen for tau air-showers (as well as for PeVs
anti-neutrino electron scattering on air electrons by resonant
interactions). Crown array detectors for horizontal Cherenkov
signals on mountains, on balloons and satellites widening the
solid angle view are being elaborated; deep and wide valleys are
considered. Tau Air-Showers neutrinos at EeV energies might rise
in AUGER, facing the Ande shadows and eventually linking twin
fluorescence telescopes to better test horizontal-inclined
Cerenkov blazing photons. Tau air-showers may be revealed in ARGO
being located within a deep valley testing inclined showers from
the mountain sides; MILAGRO (and MILAGRITO) on a mountain top
might already hide records of horizontal up-going muon bundles due
to far UHECR and less far (but rarer) tau air-showers at EeV.
MAGIC (or Veritas and Shalon) Telescopes pointing downward to
terrestrial grounds acts, for EeV Tau neutrino air-showers
astronomy, as a massive tens of $km^3$ water equivalent detector,
making (in a given direction) it at present the most powerful
dedicated neutrino telescope. MAGIC facing the sea edges must also
reveal mirrored downward UHECR Air-showers (Cherenkov) flashes.
Magic-crown systems may lead to largest neutrino detectors in near
future. They maybe located on top mountains, on planes or balloons
or in satellite arrays. They may be screened in deep valleys.
Finally within cosmological relic light neutrino mass bounds (
suggesting $\Sigma m_i\simeq 0.18 eV$)  a nearly degenerated
neutrino mass $m_i \simeq 0.06 eV$ rises offering future possible
Z-Showering  signals originated by $E_{\nu} \simeq 60 ZeV $ and
UHECR secondaries above GZK cut-off, up to  $E_p \simeq ZeV $.
\end{abstract}
%


\baselineskip=11.6pt

\section{Introduction: UHE  $\nu$ astronomy at sight}
Since Galileo we enjoyed of an optical view first of the planets,
stars,  and later galaxy maps while, since last century  we
enlarged the astronomical electromagnetic windows  in radio,
infrared, UV, X, $\gamma$  with great success. Now a more
compelling UHE $\nu$ astronomy at EeVs energy is waiting at the
corner. It is somehow linked to a very expected new particle
astronomy: the UHECR at GZK energy $\geq 4\cdot10^{19}$eV: it must
be a limited and nearby one (tens Mpc) because of cosmic BBR
opacity. There have been since now two successfully neutrino
astronomy at opposite low energy windows: the solar and the
supernova ones. The solar ones has been explained by Davis,Gallex,
SK, SNO experiment in last four decades opening the $\nu$ physics
to a solar neutrino mass splitting and a clear probe to its mixing
behavior. The supernova SN 1987A was an unique event that anyway
had a particular expected signatures at tens MeV. On going
experiment on cosmic supernova background in S.K. are at the
threshold edges, possibly ready to a
 discover of this cosmic background. However there is a more exciting and
energetic $\nu$ astronomy at PeV and EeV energy associate to the
evidence of charged UHECR spectra at EeV and tens hundred of EeV
band. Indeed any EeV CR originated nearby an AGN or GRB or BL Lac
jet will be partially screened by the same source lights leading
to a consequent photo-pion production, associated with PeV
secondary neutrinos. In a much simpler and guaranteed way, at
energy about $4\cdot10^{19}$eV, UHECR should propagate in cosmic
photon black body, being partially arrested  by  photopion
productions, (GZK cut-off), leading to EeV neutrinos all along the
Universe confines. These UHE $\nu$ components, consequence of the
GZK cut-off, are called cosmogenic or GZK neutrinos. Their flux
may be estimated by general arguments and there is quite a wide
consensus on such neutrino GZK flux at EeV energies. These
\textit{guaranteed} neutrinos may be complementary to possible
\textit{expected} higher energy neutrinos (at ZeV energies) whose
role might explain UHECR isotropy and homogeneity being originated
at cosmic distances. In this model UHECR   born as nucleons via
$\nu+\bar{\nu_R}\rightarrow Z\rightarrow X+N$ (Z-burst or Z-shower
model)
\cite{Fargion-Mele-Salis99},\cite{Weiler97},\cite{Yoshida1998} are
overcoming present (AGASA,HIRES,AUGER) un-observed local
(VIRGO,PERSEUS) source distribution, as would be prescribed by
naive GZK cut-off. However GZK neutrinos and Z-Burst neutrinos at
EeV are making comparable flux predictions and we shall restrict
to the simplest GZK flux assumption. In conclusion we remind the
role of neutrino masses in calibrating the Z-Burst showering and
the influence in UHECR spectra.

\section{Tau air showers connection to neutrino mass and
mixing} The tau production is limited,in general,to high energy
charmed mesons,whose productions are rare and severely suppressed
respect to lower energy pions ones. For this reason
$\nu_{\mu}$,$\bar{\nu_{\mu}}$ astronomy had a major attention in
last century, also for the deeper $\mu^+$,$\mu^-$ penetration with
respect $e^+$,$e^-$ and unstable tau. However the definite
$\nu_{\mu}\leftrightarrow\nu_{\tau}$ (SK data) disappearance and
the flavour neutrino mixing has given to
$\nu_{\tau}$,$\bar{\nu_{\tau}}$ a new life and attention. Indeed
the additional possibility to oscillate, even at highest energy
($10^{19}$eV) energy and  lowest mass splitting ($\Delta
m\simeq10^{-2}$eV),  is guaranteed by the huge stellar galactic
and cosmic distance ($\gg$hundred pc).
\begin{equation}\label{distanza}
  L_{\nu_{\mu}\rightarrow \nu_{\tau}}=8.3pc\biggl(\frac{E_{\nu}}{10^{19}eV}\biggr)
  \biggl({\frac{\Delta m_{ij}^2}{10^{-2}eV ^{2}}}\biggr)^{-1}
\end{equation}
respect to the above oscillatory one. In some sense this $\tau$
neutrino astronomy offers additional proof of $\nu$ mixing.It
should be noticed that on principle $\nu_{\mu}\rightarrow
\nu_{\tau}$ appearance may (or is going to) be revealed in SK
events.However the conjure of $\tau$ large threshold energy (4GeV)
and the small Earth radius size make this possibility a different
or marginal one.On the contrary a solar flare neutrino $\nu_{\mu}$
may travel and reach the Earth at threshold tens GeV energy and
convert itself successfully into $\tau$ leading to a possible
$\nu_{\tau}$ neutrino astronomy from solar flare \cite{FM}.

\section{$UHE\,\,\,\bar{\nu_{e}} + e\longrightarrow W^- $, $\nu_{\tau} \rightarrow {\tau}$, $\bar{\nu_{\tau}} \rightarrow \bar{{\tau}}$, in air?}

As the Z boson peak  favors UHE neutrinos in Z-shower model for
light neutrino masses ($E_{\bar{\nu_{e}}}\simeq
m_Z^2/2m_{\nu}\simeq 10 $ ZeV $\frac{0.4 eV}{m_{\nu}}$), in the
same way $\bar{\nu_{e}}e\rightarrow W^-$ resonance
($E_{\bar{\nu_{e}}}\simeq m_W^2/2m_e\simeq 6.3 PeV$) favors
energetic $E_{\bar{{\nu}_e}}$ hitting and showering beyond
mountain barrier (as well as within air horizontal edges). The
advantage of a mountain lay is double:a sharp filter for all the
horizontal hadronic air shower (and even muon tails) and a dense
beam dump where $\bar{\nu_{e}}e\rightarrow W^-$ or
$\nu_{\tau},(\bar{\nu_{\tau}})+N\rightarrow \tau (\bar{\tau})+X$
may take place.These events are double (first $\nu N$ or
$\bar{\nu_{e}}e$ event and later a $\tau$ decay);in water the
phenomenon has been noticed nearly ten years ago , see
\cite{Learned Pakvasa 1995}. The idea of this $\tau$ showering in
water was been considered as the double bang signatures, rarely
observable in km$^3$ detector. The some double bang reformulated
\textit{in} and \textit{out}, \textit{in} rock mountains (or
Earth) first and \textit{out} within air,later, was the main
proposal  discussed first  at the end of the previous century
\cite{Fargion1999}. In particulary since  six years ago , see
\cite{Fargion 2002a}, the upgoing and horizontal air showers has
been widely formulated for detention beyond mountain chain, as the
Alps and Ande ones. These ideas had been promptly considered for
on going AUGER experiment, just nearby Ande mountain chain, see
\cite{Fargion 2002a}\cite{Fargion2004},\cite{Miele et. all05}.
 Later on the same idea of \textit{old}
and \textit{regenerated} horizontal air shower have been
considered by other authors \cite{Bertou2002} as well as by a wide
list of additional authors \cite{Feng2002},\cite{Tseng03},
\cite{Jones04},\cite{Yoshida2004}. The difference of the $\tau$
role in its crossing the Earth lay is in its complex energy loss
processes. Ionization, bremsstralhung, pair production and
photo-nuclear losses are suppressing the $\tau$ primary energy in
such way its own lifetime may be shortened  suppressing its
propagation length. The understanding of the correct interaction
length has been noticed by \cite{Fargion1999} and it has been
incorporated on 2000 by \cite{Fargion 2002a}, and in the complete
final $\tau$ radiation length \cite{Fargion
2002a},\cite{Fargion2004}. While first and late attempts assumed a
fixed $\beta$ parameter or a linear ones, the $\beta$ dependence
with its logarithmic growth with energy has been considered
correctly by \cite{Fargion 2002a} (and not other authors) as it
has been  probed in detail only recently \cite{D05}.

\section{UHE  $\tau$   versus muon lengths}
One of the most common place in $\nu$ telescope astronomy is to
consider $\mu$ because more penetrating than $e$ and $\tau$
\cite{Fargion 2002a}. This is true in the TeV-PeV energy. However
the PeV $\tau$ is already
 to escape a mountain, decay in
flight and amplify its shower loudly, respect to a single $\mu$
escaping at some energy from a mountain. Moreover the muon
logarithmic growth  is reached at EeVs by a linear growth of an
UHE $\tau$ , mostly because if the lepton is heavier, its
electromagnetic loss is smaller. Un-fortunately hadronic losses do
not allow the $\tau$ to increase its penetration but $\tau$ is
more penetrating a those UHE regime where $\nu$ astronomy overcome
the atmosphere $\nu$ noise, see \cite{Fargion 2002a}. In more
sophisticated approach, not shown here for sake of simplicity, one
may estimate the Earth skin to Tau Air-shower for $shorter$
maximal lengths that guarantee a unsuppressed $highest$  Tau
escape energy; this minimal Earth skin define a smaller volume and
lower tau air-showering rate, but at highest $EeVs$ energy
\cite{Fargion2004}.

\begin{figure}[b!]
\vspace{7.0cm}
 \includegraphics{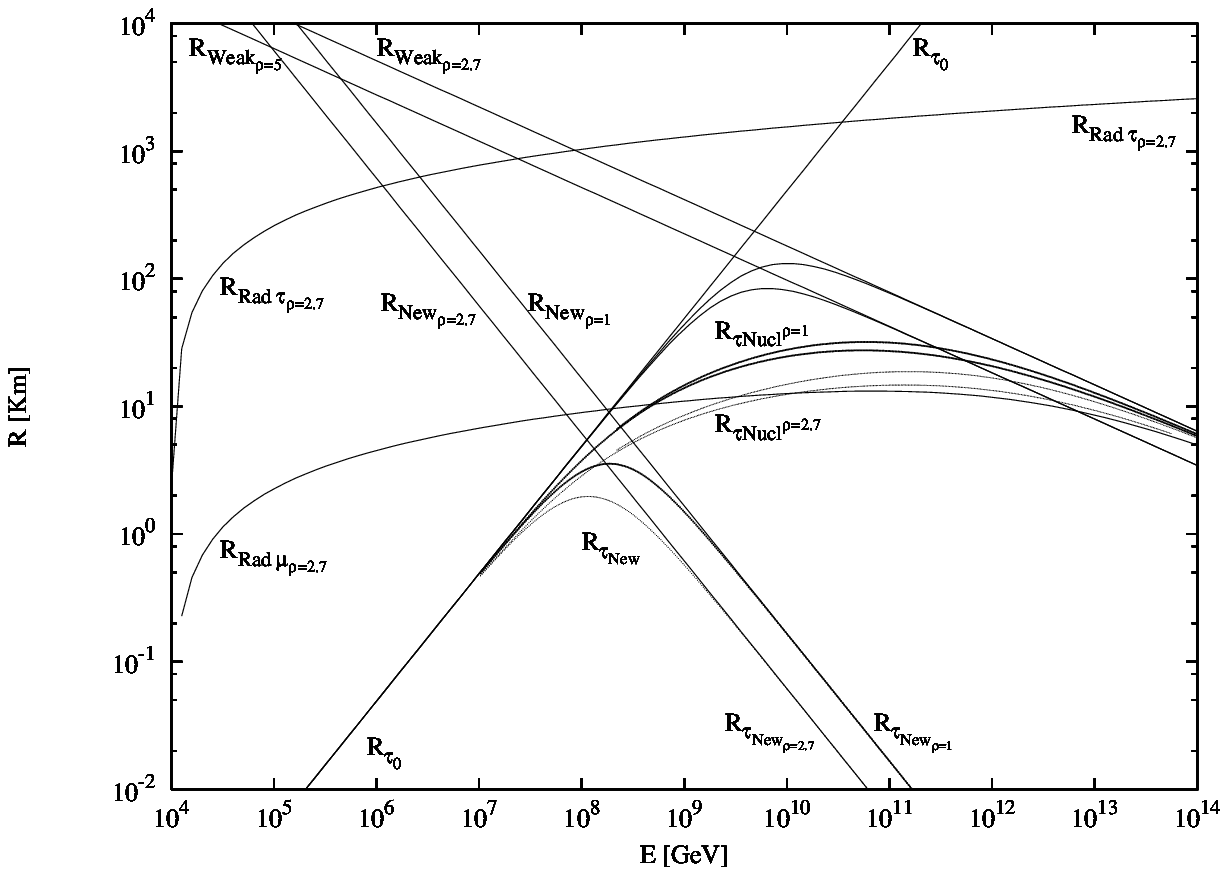}
\vspace{-1.0cm}
 \caption {\it{While tau are extremely unstable and electron
are leading to short radiation length, muons are usually the most
penetrating lepton; this usually favor underground muon neutrino
telescope. However at ultra-relativistic regime tau reach and
overcome the muon tracks, making the heaviest lepton the most
penetrating. Because life-time linkage to tau energy and to its
energy losses,dominated by hadron and pair production, the tau
interaction length is derived by an hybrid transcendent equation
comparing energy losses and life-time length.  }}
\label{fig:fig1}
\end{figure}

\begin{figure}[b!]
\vspace{8.0cm}
\includegraphics{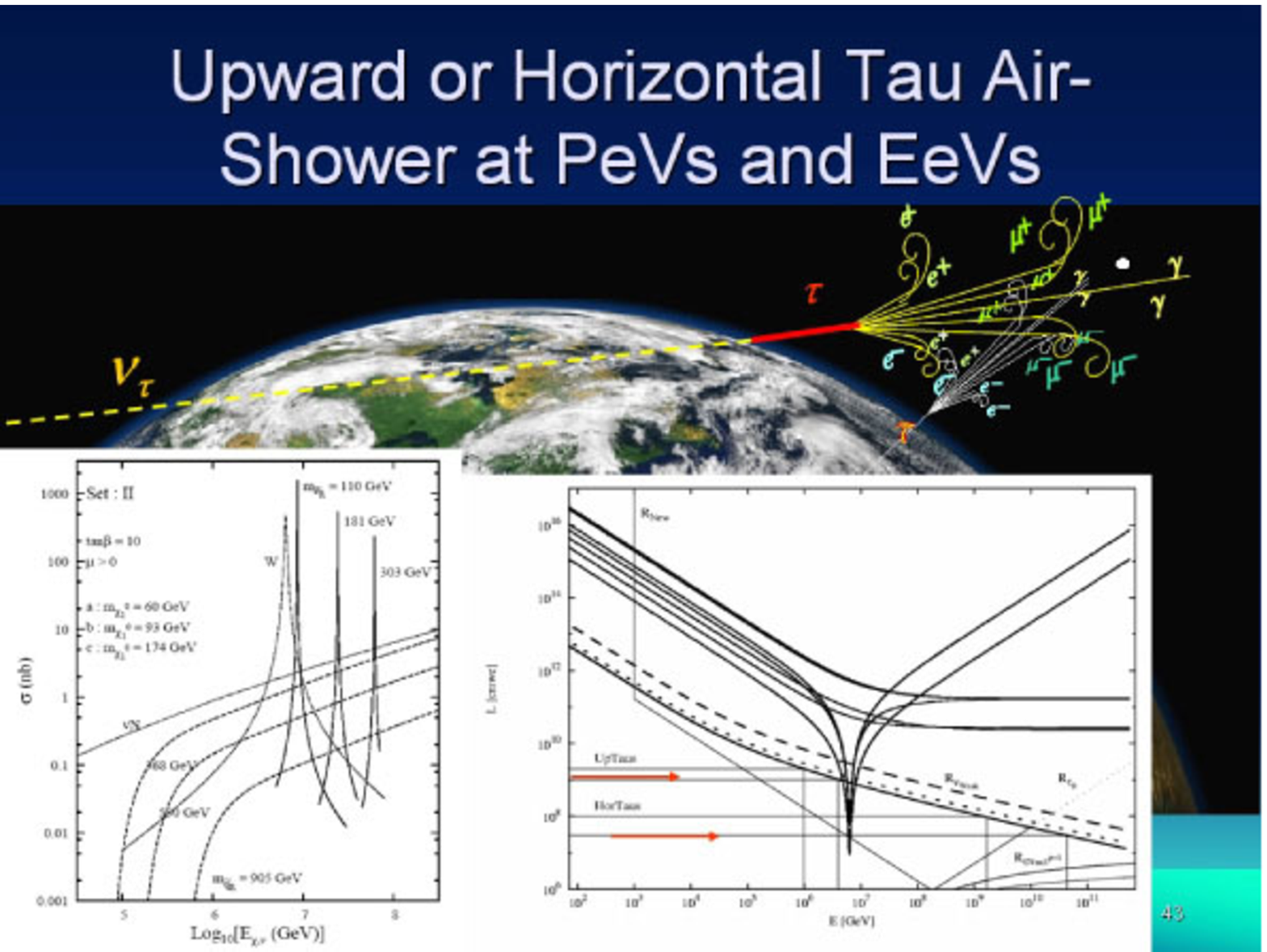}
\caption {{\it A schematic view of the possible Horizontal
Tau Air-showers at EeVs energy versus a lower PeV vertical
up-going Tau Air-Shower. In the left side insert the
cross-sections for UHE anti-neutrino with electrons, mediated  by
$W^-$ and the almost comparable UHE neutralino scattering on
electron leading to $\tilde{e_R}$  whose decay in flight lead also
to UHE electromagnetic jet-shower. The consequent interaction
length for both neutrino with nucleons and its peculiar
anti-neutrino-electron  Glashow resonance is shown in the second
insert. The Earth diameter in nearly $10^{10}$ cm. water
equivalent; therefore  the terrestrial neutrino opacity arises
above tens PeV energy or inside the narrow  resonant Glashow's
neutrino peak. To overcome this neutrino opacity one may consider
mountain chains or small (PeV) Uptaus or shorter terrestrial cord,
for higher energy Hortaus,that are just at the horizons as shown
by the red arrows } see \cite{Datta}.\label{fig:fig1}}
\end{figure}

The behavior of $\tau$ lengths for most adopted energy losses,the
possible definition of a shorter length that guarantee a higher
outgoing $\tau$ energy,all the detailed Earth profile density for
escaping $\tau$ and the consideration of the finite atmosphere
size for escaping $\tau$ air shower, all these details have been
analyzed in a tail of recent article \cite{F04}.Independent
attention has grown in studying the upgoing $\tau$ flux in km$^3$
detectors \cite{Feng2002},\cite{Tseng03},
\cite{Jones04},\cite{Yoshida2004}. The general results are not
always converging and a summary of the most recently results has
been shown (see last fig  in ref.\cite{Fargion2004} for general
comparision).
comparision).
 While we have not yet experienced $\tau$ definitive
air showers , we may foresee that any neutrino $\tau$ astronomy
will, soon or later, test the $\tau$ decay channels. Indeed the
main multiple $\tau$ decay channel are leading to weighted channel
and showers described in PDG table. It will be possible,in
principle,to verify by ratio of $\bar{\nu_e}e\rightarrow
W\rightarrow \tau$ \textit{monocromatic} channel versus
$\nu_{\tau}+N\rightarrow \tau$ channel, the $\nu_{\tau}/\nu_e$
abundance and the primary flavour mixing. In a few words $\tau$
air shower must be consistent and correlated, in its decay mode by
electromagnetic, hadronic and hybrid channel, with well known
elementary particle result.

\section{Many Roads to Tau Air-showers}
  There are very advanced experiment that ( wherever they are aware or not) might
  point for  Tau Air-Showers, even they were originally thought for
  other scientific targets.  The list of these High energy Showering experiment
     adaptable to tau Neutrino and Horizontal Showering is here briefly reviewed:
    1)Argo, in Tibet; 2) Milagro (and Milagrito) in USA mountains,3) AUGER experiment in Argentina, 4) Space
    Station Crown Arrays (to be effectively proposed \cite{FarCrown}), 5) EUSO telescope, 6)  BATSE satellite in CGRO ($1991-2000$ past),
    7) ASHRA experiment in Hawaii (now continued by NuTel collaboration), 8) CRTNT Fluorescence array in Utah or China,\cite{Cao} 9)
    Muon Array Telescope in Jungfraujoch,\cite{Iori04} 10)  Cherenkov
    Telescopes on High Mountains facing the mountains, like Shalon in Kazakistan.
    In this view the Magic Stereo Telescopes facing the Earth edges are somehow ideal \cite{Fargion2005}.
     The first $Horizons$ tests maybe done possibly in cloudy and otherwise astronomical useless
     nights.   Here below the images and the
  captions explaining how those experiments may find Tau Air-Showering by a minimal optimized
  trigger set up.
  \subsection{ARGO}
  This large area array inside  a deep valley in Tibet may record
  PeVs Tau air-showers emerging from the mountains around. The
  nearby Chines-Japanese twin experiment may enlarge the area. The
  presence of more (tens-hundreds) spread (small, few $m^2$ area) elements at hundred
  meters one from the other, in vertical structure as well as the
  covering of the inner wall periphery of the detector house, may
  greatly increase the ARGO ability to reveal PeVs air-showering
  below the mountain shapes. The  variable opacity to atmospheric
  GeVs-TeVs muons within the mountain shadows,  is a needed test.
\begin{figure}[b!]
\vspace{5.0cm} \includegraphics{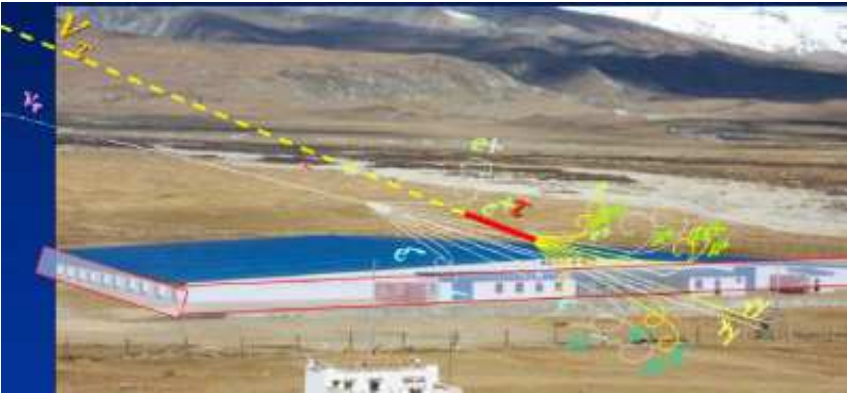}
\caption {{\it The possible use of Italian-Chinese ARGO (and its
twin nearby Chinese-Japanese array) to monitor, inside a wide deep
valley, inclined or horizontal tau air-showers originated by
surrounding mountains; the signal may be better revealed by
additional array detectors on the walls along the lateral
boundaries; these lateral-wall array are  in analogy to  present
Nevod-Decor detector parallelepiped  structures, in
Russia.}\label{fig:fig1}}
\end{figure}
\begin{figure}[b!]
\vspace{6.5cm} \includegraphics{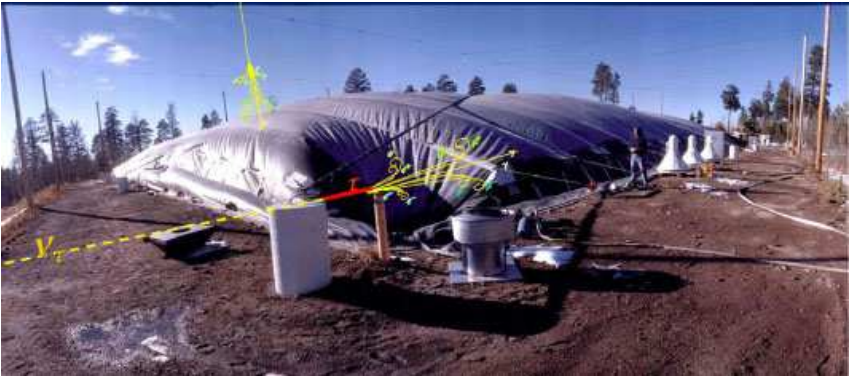}
\caption  {{\it The possible UHECR horizontal or up-going Tau
air-showering on Milagro (as well in correlated mode, to nearby
Milagrito) while being  a TeV gamma detectors: GRB or an active BL
Lac at horizons, making nearly $1-3\%$ of GRB,SGRs,BL Lac events ,
might play a role in shining and tracing muon bundles in the
Milagro pool waters. As a first estimate, assuming an effective
area of few $10^3 \cdot m^2$ we foresee one or a few events of
Upward muon bundles associated to Tau Air-Showers each year,
depending on the trigger, the  threshold and geometry.
}\label{fig:fig1}}
\end{figure}
  \subsection{MILAGRO}
   The existence of huge pools at peak mountains as Milagro and
   smaller Milagrito, offer an exiting laboratory to verify
   (besides TeV gamma backgrounds): the muon horizontal fluxes at
   horizons; the muon bundle density, flux and structures (in
   comparison with sea-level Nemo-Decor data, see \cite{Decor},\cite{NEVOD}); finally there is
   the possibility to discover Up-going muon bundles, whose
   existence maybe indebt only to Earth Skimming (Uptaus)
   Air-showering.
 \subsection{AUGER}
 As in figures and in captions the Auger experiment offer a unique
  occasion to Horizontal Tau possibly from the West side toward
  the AUGER detector; to optimize the ability to disentangle these
  events one should first observe the Ande shadows (at $87-90^o$),
  zenith angles by a simple asymmetry East-West UHECR showering,
  see\cite{Fargion1999},\cite{Fargion 2002a}\cite{Fargion2004},\cite{Miele et. all05}.
  Within the first year of full operation the shadow $must$ be
  seen. Later on, within the same solid angle of $\simeq 2\cdot 100 = 6\cdot 10^{-2} sr.$
 two event a year by tau Air-showers (via GZK neutrino flux)
 $might$ be  very probably observed see \cite{Fargion2004}.

\begin{figure}[b!]
\vspace{5.0cm} \includegraphics{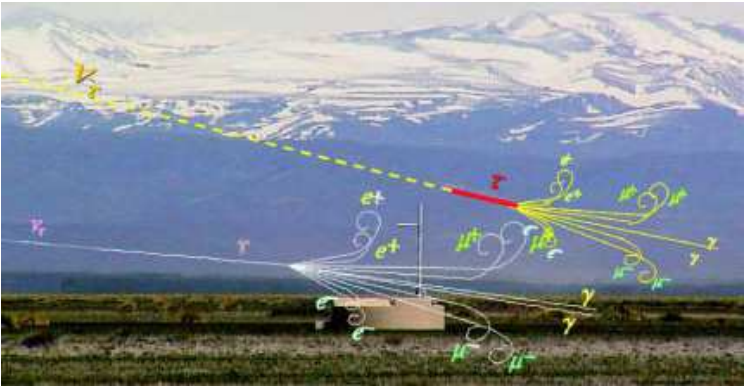}
\caption {{\it The long Ande chain mountain is offering a unique
wide screening shadows (for UHECR) (opening angle $2-3^o$) and an
ideal beam dump (for PeVs-EeVs tau neutrinos) to AUGER array
detector. Inside this shadows, that may be soon  manifest, rare (a
few a year), but quite guaranteed horizontal tau (by GZK neutrino
fluxes) air-shower that might be open Neutrino Astronomy
windows.}} \label{fig:fig1}
\end{figure}

\begin{figure}[b!]
\vspace{6.0cm} \includegraphics{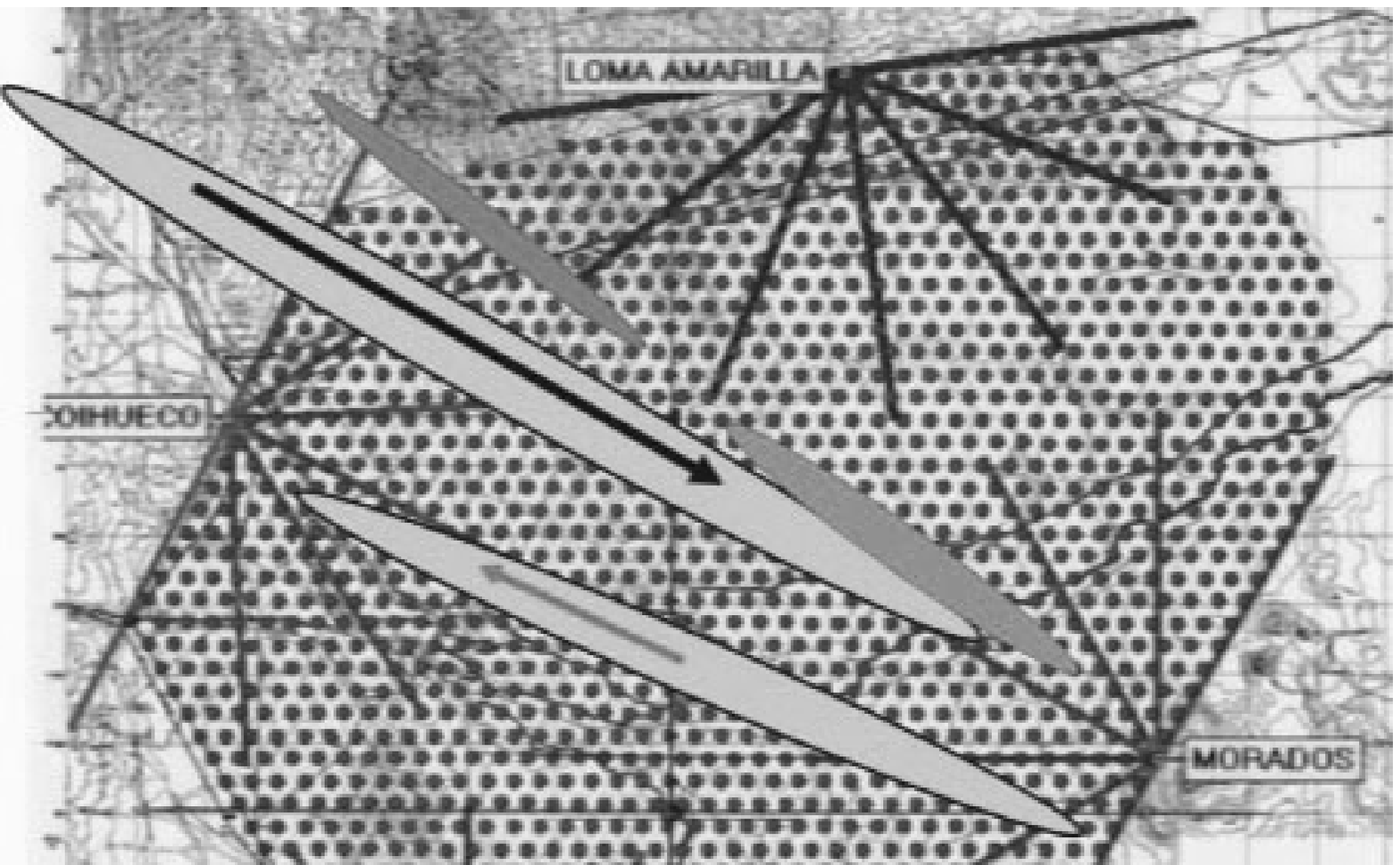}
\caption  {{\it Inclined Horizontal Air-Showers able to trigger
$both$ Auger tanks $and$ Fluorescence telescopes , while being in
the same axis. This tecnique, as long the author knows, has never
been used to better disentangle horizontal Air-Showers. It may be
an ideal detector to observe Tau Air-Showers from the Ande. Their
events will populate the forbidden area of large zenith angle at
horizons. For this reason it will be useful to: a) enlarge the
angle of view of Coiuheco (as well as Loma Amarilla and Leones
station) toward the Ande;
 b) to eliminate any optical filter for Cherenkov lights in those directions ;
 c) to open a trigger between the Array-Telescope, or Telescope-Telescope in Cherenkov blazing mode; d) to try all $4$
telescope Fluorescence connection in Cherenkov common trigger-mode
 along all the $6\cdot 2 = 12$ common arrival directions.  Similar
connection along the $360^o$ view of stereoscopic HIRES
telescopes, might be already done testing along  their common
horizontal $2$ axis   air-showers (from PeVs up to EeV energy)
with  high rate (tens-hundreds events a night)
 and great angular accuracy. In the picture some possible inclined UHECR
 events shining both array detectors and  (by Cherenkov lights) Fluorescence Station;
  possible twin separated ovals arise by geomagnetic bending.}}
  \label{fig:fig1}
\end{figure}

    The AUGER Fluorescence detector may enlarge their view also
  toward the Ande, offering an ideal screen capturing Ande-Tau
  Showers in horizontal tracks at best. The possibility to use
  inclined air-shower Cherenkov lights hitting the Fluorescence
  detector maybe exploited. Multi-telescope coincident Cherenkov
  detection (while being nearly on axis) of horizontal air-showers
  maybe applied in all the $12$ common directions ($4\cdot3$) in AUGER (and $2$ for
  stereoscopic HIRES).
 \subsection{CROWN ARRAYS ON SPACE STATION }
\begin{figure}[b!]
\vspace{5.0cm} \includegraphics{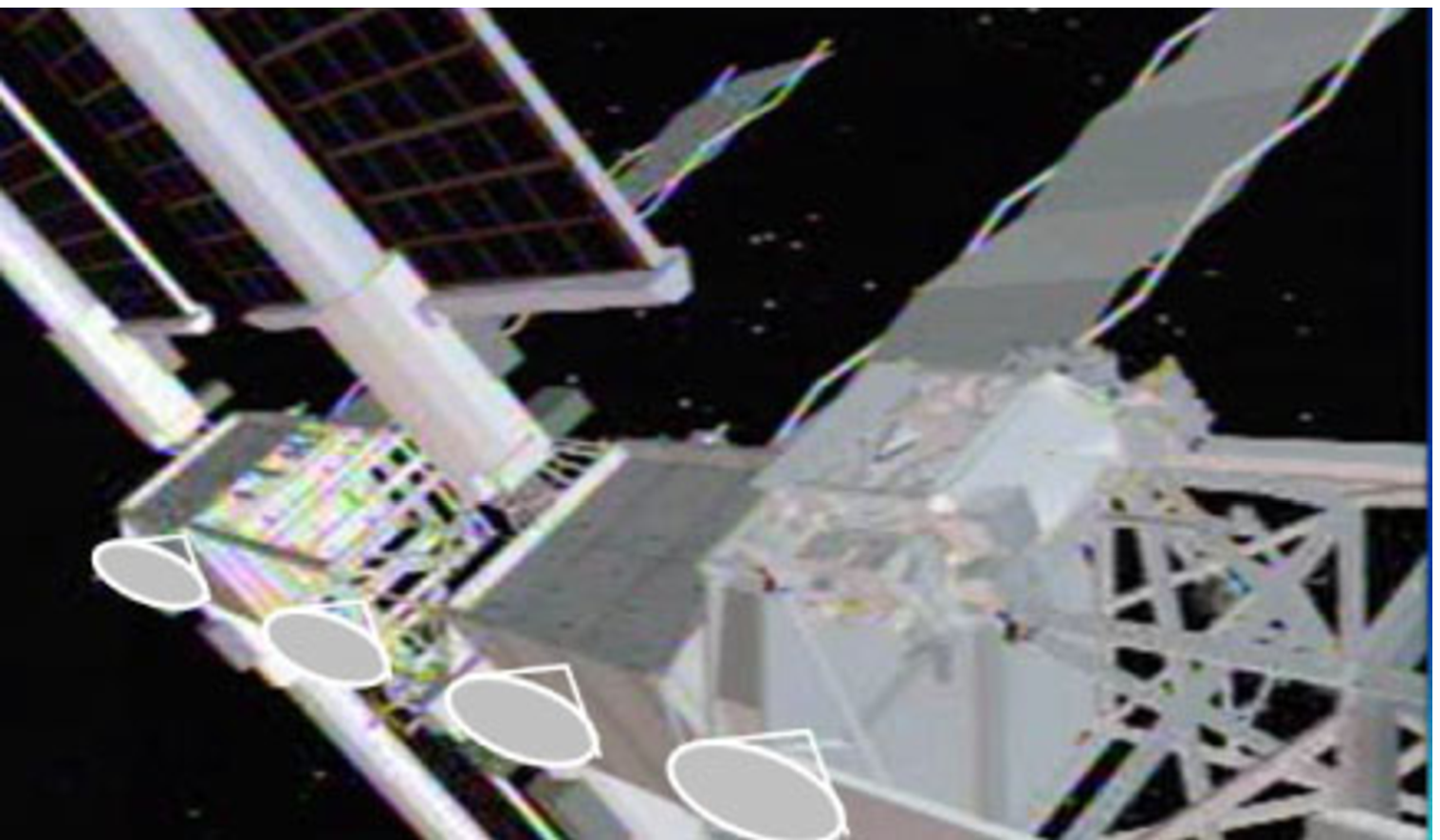}
\caption  {{\it Space Station constructed and armed with a
Telescope Cherenkov Array and a Gamma Array spread array able to
disentangle gamma flashes and arrival lights, from the Earth
edges. The possible Tau neutrino nature is imprinted by the
arrival direction $below $ the Earth horizons, while the UHECR
showers arise at the high Atmosphere (Albedo) edges above. The
duration of the signal (micro-second to millisecond), as for
Terrestrial Gamma Flashes, is the signature of these Up-going
Air-Showers, steady ones are the signature of Gamma TeVs-PeVs
air-showering sources. The threshold depends on the Telescope and
Gamma detector areas; even the distances from Space Station are
nearly $100-200$ larger than vertical TeV-PeV air-showers on
Earth, the beaming is $14-20$ smaller, with negligible absorption,
making a $2$ meter square Cherenkov telescope able  in principle
to observe TeVs gamma sources.) }See \cite{FarCrown}}
\label{fig:fig1}
\end{figure}
  From the Space there is the most appealing location to search
  for Horizontal High Altitudes Showers and Horizontal or Vertical
  Tau-Air-showers (Hortau-Uptau). This project is still
  preliminary. The crown-array maybe $both$ detecting (tens, hundred keV) gamma secondaries
 (as well as rarer hundred GeV lepton pairs)
  as well as Cherenkov lights due to far Hadron and Gamma primary High
  Energy Cosmic Rays showering from Earth.  The array maybe at
  PeVs-EeVs energy equivalent to few-hundred km. mass neutrino
  detectors, depending on the telescope sizes and gamma array
  area.See \cite{Fargion2001},\cite{FarCrown}.
\subsection{EUSO }
The project of a telescope facing down-ward the Earth and catching
the UHECR has been delayed to the end of the century. However the
idea may offer a way to discover beamed horizontal HorTaus at tens
EeV energy showering at high altitudes. Few events , $4-6$, might
be observed each year. The EUSO mass equivalent due to
Earth-Skimming  \cite{Feng2002}, or HorTaus \cite{Fargion 2002a}
neutrinos is nearly $100 sr km^3$ \cite{Fargion2004} water
equivalent, even taking into account the $10\%$ duty cycle of the
EUSO activity.See \cite{Fargion2004}.

\begin{figure}[b!]
\vspace{6.0cm} \includegraphics{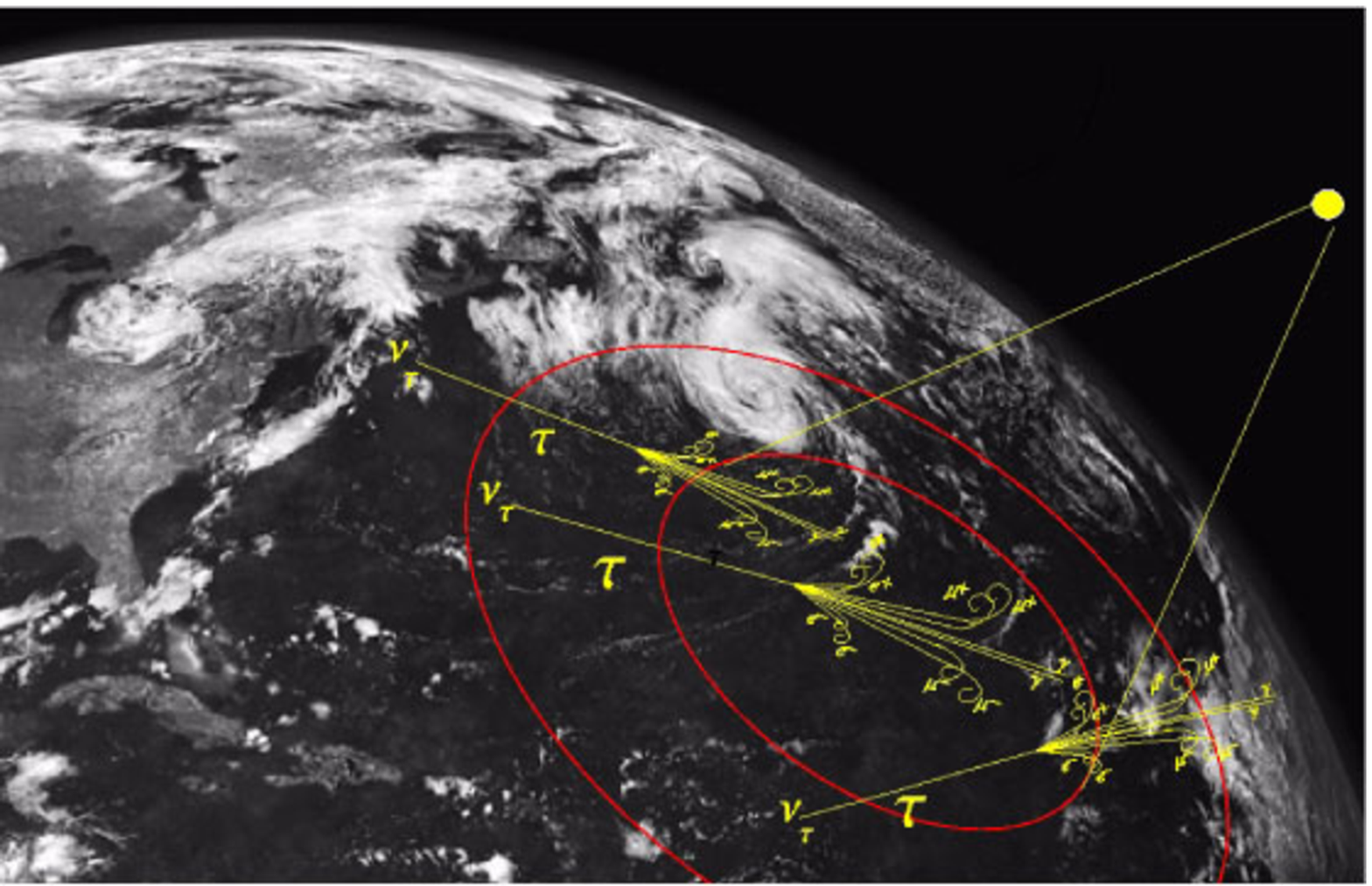}
 \caption  {{\it The up-going horizontal air-showers whose longest (hundreds of km.) air-showers
 might be detectable by future Euso project; the project would reveal thousands oh UHECR
 mostly downward events, as well as hundred of horizontal C.R.
 airs-showers, whose beam angle is extremely small, because low air density.
 Within these down-going UHECR air-showers there are 4-6 event a year originated
 within a wider field of view. One of his greatest proponent and great
 scientist,that with Prof. Linsley discovered   UHECR  at GZK  edges,
 Prof. Livio Scarsi, sadly has very recently missed.} See  \cite{Fargion2004}}
 \label{fig:fig1}
\end{figure}


\begin{figure}[b!]
\vspace{8.0cm} \includegraphics{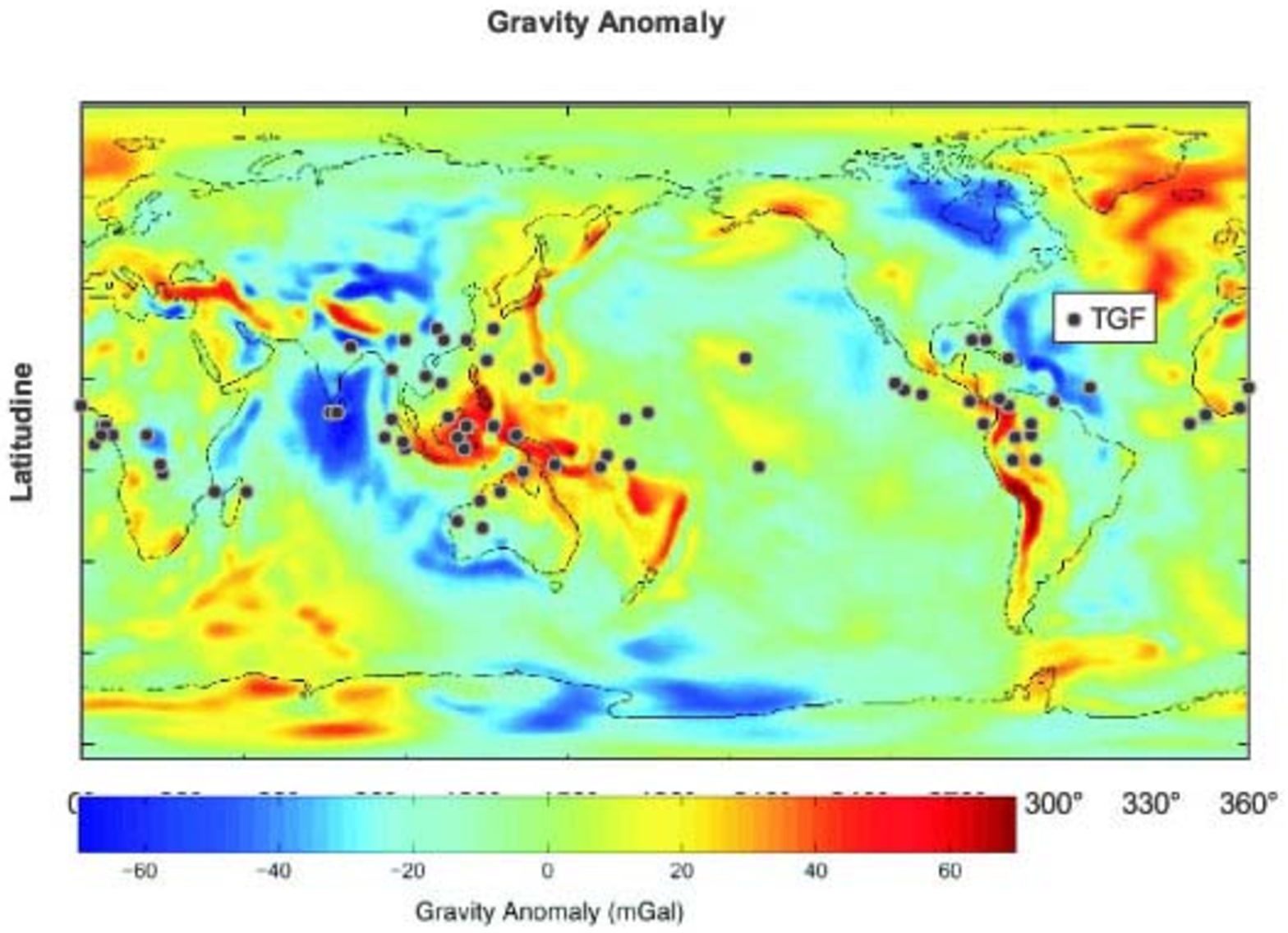}
 \caption  {{\it The apparent correlations between Earth
crust contrast, gravity anomaly and the observed location of
Terrestrial Gamma Flashes observed by BATSE in last decade (and by
RHESSI last two years). The overlap of the TGF events with maximal
terrestrial mass density contrast (Mountain chains, sea-islands)
(in the equatorial belt where BATSE-Compton trajectory laid),
favors a common origin of TGF and tau-airshowers. }See
\cite{Fargion 2002a}, \cite{Fargion03}} \label{fig:fig1}
\end{figure}

\subsection{EOLIC ARRAY}
 On the top of the mountains small crown  arrays of detectors may be
 mounted on the eolian energy stands.
 The usual sites ( mountain)
 the power supply, the two different hight on the same element may
 offer a useful place to build an horizontal Air-shower detector,
 in wild areas and spread surfaces \cite{FarCrown}.

\begin{figure}[b!]
\vspace{6.0cm} \includegraphics{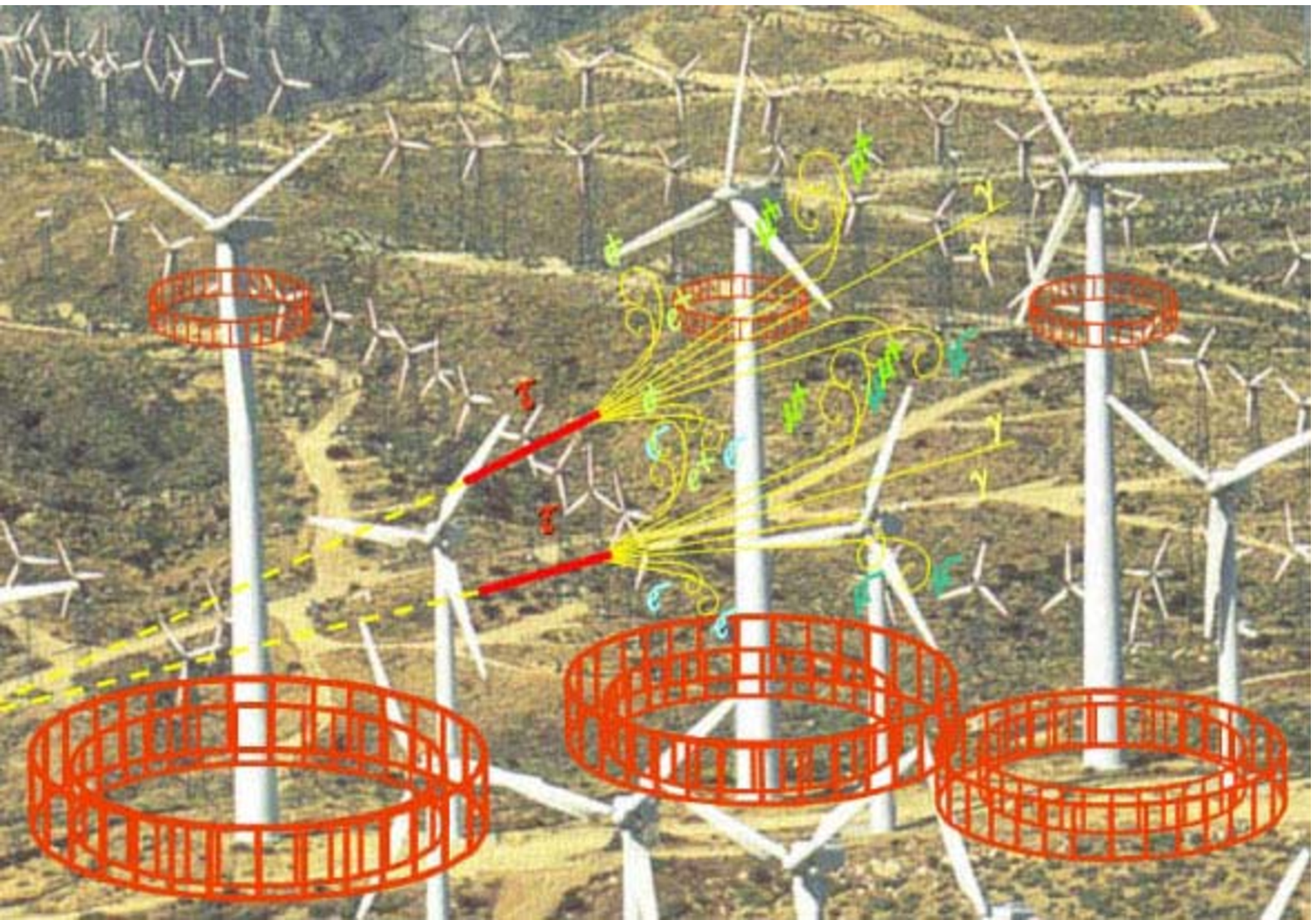}
 \caption  {{\it Ideal arrays of crown scintillators on wind eolic stations.}}
  \label{fig:fig1}
\end{figure}

\subsection{BATSE}
Old generation of Gamma satellite in orbit last decade made (with
deep discovers by Beppo Sax) most of our view in gamma astronomy.
Present and next generation (Swift,Glast) will enlarge the EGRET
astronomy by deeper views. The same skimming C.R. or Albedo and
Air-Shower tracing by UHECR (PeVs-EeVs) will naturally arise
\cite{Fargion2001}.
\subsection{ASHRA }
Three  Fluorescence detectors, in a similar way as
  AUGER telescopes, are monitoring from the top mountains of the Hawaii island the
  inner area; their detection maybe  greatly enhanced by tracing and calibrating higher altitude HIAS and
  facing the Earth edges, searching the HorTaus at ocean Horizons.



\subsection{JUNGFRAUJOCH} The existence in the top Europe turistic station
of scientific facilities and fast transport made possible a first
test of  Muon Telescope Array prototype at horizons site. The
proposal of a larger area and more numerous detector is in
progress and it may compete with Cherenkov telescope also because
of light noise independence of the scintillator array.
\subsection{CRNT}
The proposal of a Fluorescence array within the cliff shadows is
going to be considered in Utah and-or in China. The PeV detection
will be possible by low noise light location and Cherenkov  aided
discovering techniques.
 \subsection{SHALON}
 A Russian proposal leaded by Cherenkov TeV-Telescope is already
 looking  from  the mountains terrestrial
 targets in search of eventual Tau air-showers, finding already a statistics
 on Albedo air-Showers and relevant first calibrations.

\section{The Veritas and  Magic views  of Tau Air-Showers at horizons}

Cherenkov gamma Telescopes as last Veritas and  MAGIC ones at the
top of a mountains are searching for tens GeV $\gamma$ astronomy.
The same telescope at zero cost in cloudy nights, may turn (for an
bending angle $\simeq 10^{\circ} $) toward terrestrial horizontal
edges, testing both common PeVs cosmic ray air showers, muon
secondary noises and bundles as well as upgoing tau air-showers.
Indeed the possible detection of a far air shower is enriched by:
\begin{enumerate}
  \item early Cerenkov flash even dimmed by atmosphere screen
  \item single and multiple muon bundle  shining   Cerenkov rings or
  arcs inside the disk  in time correlation
  \item muon decaying into electromagnetic in flight making
  mini shower mostly outside the disk leading, to lateral correlated gamma tails.

\end{enumerate}
  We estimated the rate for such PeVs-EeV events each night,
  finding hundreds event of noises muons and tens of bundle
  correlated signals each night \cite{Fargion2005}. Among them  up-going Tau
  Air-Showers may occur very rarely, but their discover is at hand for
  dedicated $360^o$ crown Arrays \cite{FarCrown}(and arrays of these crowns) in correlation among
  themselves and scintillator detectors.

\begin{figure}[b!]
\vspace{3.0cm} \includegraphics{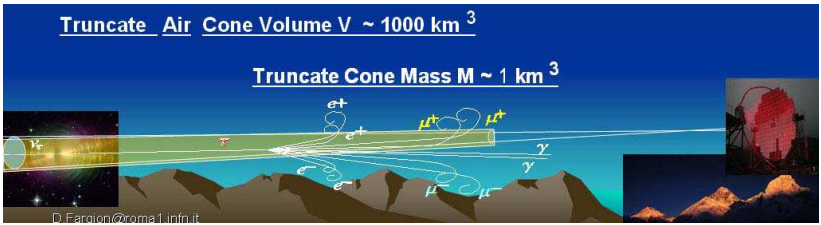}
\caption  {{\it The possible horizontal air-showering by a GRB or
an active BL Lac , whose UHE anti-electron neutrino might
resonance with air electrons at Glashow PeVs energies (or in
 Tau air-showers at higher energy), making  nearly $3\%$ of these GRB,SGRs,BL Lac
 sources  laying at horizons for Magic  Telescopes. The mass observed , as estimated in figure, within the
 air-cone exceed the $km^3$ water mass, even if within a narrow solid angle ($\simeq 4\cdot 10^{-3}$ sr.) }See \cite{Fargion2005}}.
 \label{fig:fig1}
\end{figure}
\begin{figure}[b!]
\vspace{3.0cm} \includegraphics{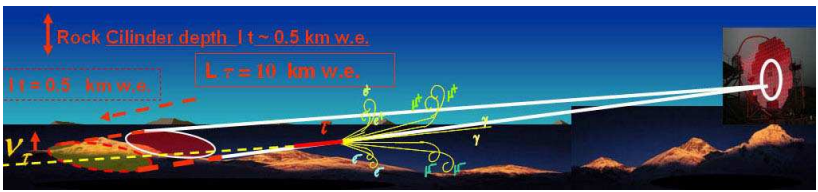}
 \caption  {{\it As above  EeVs tau  are originated in the Earth crust
 and while escaping the soil are testing $\sim 70-100 km^3$ volumes; later
 UHE tau may decay in flight and may air-shower loudly toward Magic telescope, within an area of
 few or tens $km^2$.} See \cite{Fargion2005}.}
 \label{fig:fig1}
 \end{figure}
\begin{figure}[b!]
\vspace{6.0cm} \includegraphics{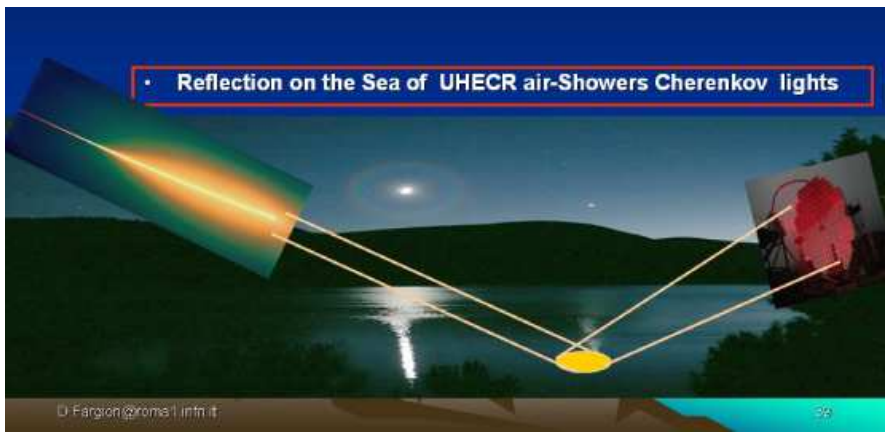}
 \caption  {{\it The possible inclined UHECR air-showering on
Magic facing the sea side. Their detection rate is large (at
zenith angle $80-85^o$) (tens or more a night)
 nearly comparable with those  at zenith angle  $87^o$ already estimated;
  these mirror UHECR shower , widely spread in oval
  images on the sea (depending on the sea wave surfaces), their presence
is an useful test for Magic discovering of point source PeV-EeV
UHECR air-showers at horizons.
 While previous configuration above horizons may correlate direct muon bundle and
Cherenkov flashes, these mirror events are polarized lights mostly
muon-free, diffused in large areas and dispersed in longer time
scales, mostly in twin (real-mirror-tail) spots. On the contrary
Up-going Tau air-showers from the sea are very beamed and thin and
un-polarized and brief.} See \cite{Fargion2005}.}
\end{figure}

\begin{figure}[b!]
\vspace{5.0cm} \includegraphics{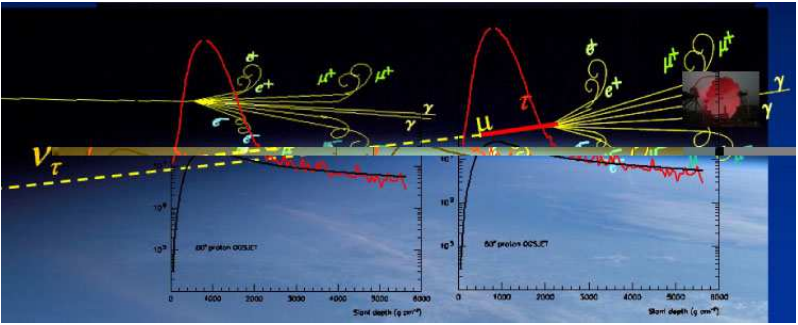} \caption  {{\it The  horizontal air-showers by far
hadron differ to an up-tau air-showers,
 whose younger electromagnetic and muonic density is greater and much larger ;
  in the figure the two different
 signature of the flux densities assuming a Magic telescope observer (not in scale),
  and an ideal downward far nucleon and a nearby Tau EeV air-showers event.} \cite{Cillis2001}}
\end{figure}

\begin{figure}[b!]
\vspace{5.0cm} \includegraphics{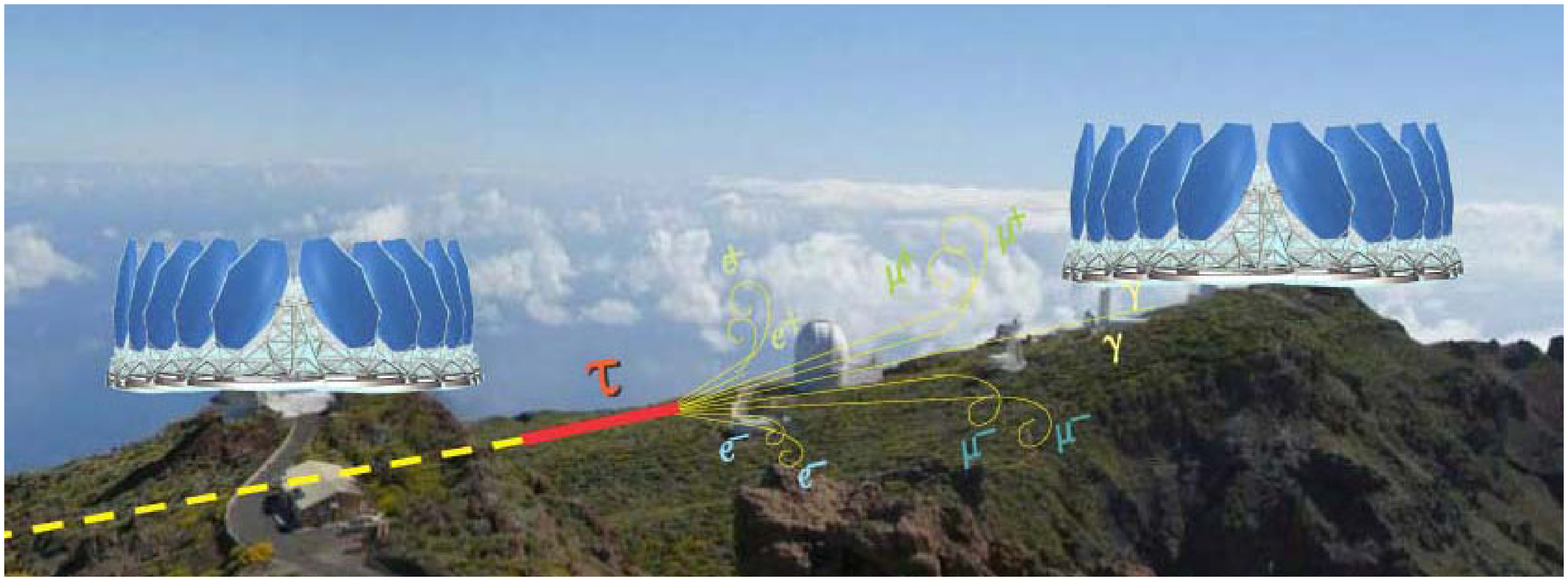}
 \caption  {{\it Ideal arrays of Cherenkov Crowns Telescopes in
 Canaries and an equivalent twin Crown Array Balloon  in flight; similar arrays maybe located
 in planes or satellites.} }
 \label{fig:fig1}
\end{figure}
The timing of these signals,their expected event rate at PeVs in a
night time of Magic at horizon ($87^o$ zenith angle ) and their
easy signature has been reported recently by Fargion
\cite{Fargion2005}.More over the very simple exercise of the
estimate of the air cone volume observed at the horizons by MAGIC
shows a value larger than $10^3Km^3$ corresponding to a mass
volume larger than $1km^3$ water equivalent.This volume at Glashow
resonance energies make MAGIC the must wide $\nu$ detector.The
some estimate even at smaller solid angle,below the horizon leads
a large mass for EeV neutrinos,encompassing volumes and masses as
large as $10^2km^3$.These detectors are active only within a
narrow view,but during peculiar rise and down of BL Lac,AGN or
Crab like sources, or in coincidence with GRBs along the
horizon,the masses enlisted are huge and relevant.To make the
detection permanent and in wider angle view the ideal crown array
of MAGIC-like telescope on circle and their twin or multiple array
structure at few km distance,will guarantee a huge capability to
observe an event or few event  of $\tau$ upgoing shower during a
month,within a few tens of UHECR above the edges.

\section{Conclusions : Neutrino Astronomy and Spectroscopy}
Because muon tracks are  mostly  of downward atmospheric nature
the underground neutrino telescope are tracing rarer upward ones
mostly of  atmospheric origin;  higher energy up-going
astrophysical neutrino signals are partially suppressed by Earth
opacity, and are unique tracks. On the contrary $\tau$ air shower
at horizons is spreading its signal in a wider area  leading to
populated (millions-billions) muon and gamma (as well as electron
pairs)  bundles in their showering secondary mode. This amplified
signal may be observed and disentangled from farer and air
filtered UHECR, in different ways and places: mountains, balloons,
satellites with different detector array area and thresholds. The
advantage to be in high quota is to be extending the visible
target terrestrial area and solid angle, as well as to let a
longer tau flight distance (and energy), and to enlarge the air
shower area; a too high altitude, however , looses solid angle. To
make an intuitive estimate the Tau air-shower size area, at tens
PeVs-EeVs ,( detectable at horizons within a lateral distance as
large as $3$ km. from the main shower axis by a telescope like
Magic),it is nearly $30 km^2$; at EeV energy the equivalent
detection depth crossed by the tau lepton before the exit from the
Earth reaches $10-20 km$ distances; the corresponding  detection
Neutrino volume (inside the narrow, conic $10^{-3} sr.$, shower
beam)   is within $30-60 km^3$, in any given direction , see
\cite{Fargion2005}.
   A few  events of GRBs  a year may be located within these horizons,
   as well as AGN and BL Lac in their flare activity. In such occasions
   Magic, Veritas and Hess array are the most sensitive neutrino telescope at PeVs-EeV energy.
  Even on average, for a present $2\cdot2^o$
 view of Magic, at present energy thresholds, such telescopes
 (for Neutrino at Glashow PeVs energy  windows), are testing
a total mass-solid angle a comparable or larger than to  $10^{-2}
km^3 sr$, an  order of magnitude comparable with the present
AMANDA detector.
     Moreover the light neutrino mass that seem to converge (by
     recent cosmic constrains) toward a light degenerated neutrino mass value
    $m_i \simeq 0.6 eV$ seem to suggest a UHE primary neutrino at $E_{\nu} =
    60 ZeV$ and a UHECR secondary bump at $E=3.3 ZeV $.
      In this view UHECR modulation at GZK energies may reflect
        lightest neutrino masses.

     In conclusion  a maximal alert for the Neutrino
     air-showering within the Earth shadows is needed: in AUGER, Milagro,
     Argo, as well as in ASHRA, CRTNT, Shalon Telescopes the signal is
     beyond the corner. In particular   the  Magic (and Veritas) arrays telescopes facing
      from the mountains the Horizons   edges may soon  test our proposal leading to such crown arrays.
      In a sentence we believe that the UHE Neutrino Astronomy
     is beyond the corner, Tau is its courier  and its sky lay just  beneath
     our own  sky: the Earth.

\end{document}